\documentclass[9pt,twocolumn,twoside]{opticajnl}
\journal{opticajournal} % use for journal or Optica Open submissions

\setboolean{shortarticle}{true}
\usepackage{bm}
\usepackage{lineno}
\usepackage{graphicx} %导入包
\title{A novel and efficient parameter estimation of the Lognormal-Rician  turbulence  model based on $k$-Nearest Neighbor and data generation method}

\author[1,*]{Maoke Miao}
\author[2]{Xinyu Zhang}
\author[3]{Bo Liu}
\author[3]{Rui Yin}
\author[3]{Jiantao Yuan}
\author[3]{Feng Gao}
\author[3]{Xiao-yu Chen}

\affil[1]{Foundation Science Education Center, Hangzhou City University, 310015, China}
\affil[2]{Department of Communications and Networking, Aalto University, Espoo 02150, Finland}
\affil[3]{School of Information and Electrical Engineering, Hangzhou City University, Hangzhou 310015, China}
\affil[*]{miaomk@hzcu.edu.cn}
\begin{abstract}
In this paper,  we propose a novel and efficient parameter estimator based on $k$-Nearest Neighbor ($k$NN) and data generation method  for the Lognormal-Rician turbulence channel, which is of vital importance to the free-space optical/quantum communications. The Kolmogorov-Smirnov  goodness-of-fit statistical tools are   employed to investigate the validity of $k$NN approximation under different channel conditions and  it is shown that the choice of $k$ plays a  significant  role in  the approximation accuracy. We present several numerical results to illustrate that solving the constructed objective function can provide a reasonable estimate of the actual values. The mean square error simulation results show that increasing the number of generated  samples by two orders of magnitude does not lead to a significant improvement in  estimation performance when solving the optimization problem by the gradient descent algorithm. However, the estimation performance under the genetic algorithm (GA) approximates to that of the saddlepoint approximation and expectation-maximization estimators. Therefore, combined with the GA, we demonstrate that the proposed estimator achieves the best tradeoff between the  computation complexity and the accuracy.
\end{abstract}

\setboolean{displaycopyright}{false} % Do not include copyright or licensing information in submission.

\begin{document}

\maketitle

\section{Introduction}
\indent Recently, free-space optical communication and free-space quantum communication have attracted great attention as free-space channel has many  advantages over  optical fiber channel. For example,  free-space channel is more conducive to long-distance optical transmission since optical power in optical fiber  decreases exponentially with distance. Besides, it is more economical and flexible to  transmit  optical signals through  free-space channel in remote areas or over challenging terrain \cite{Chai2020,Brougham2022,Jeon23}. \\
\indent However, in spite of the several advantages of a free-space channel, the atmospheric turbulence within the Earth's atmosphere can significantly deteriorate the practical communication systems. In free-space optical communication, it is acknowledged that one of the performance-limiting factors is  turbulence-induced scintillation, which contributes to the excess noise, an important parameter determining the performance of continuous-variable quantum communication \cite{Miao22,Zhang23}. Hence, to precisely evaluate communication system performance in different channel scenarios, the accurate  establishment of a scintillation model is very important. Up to now, depending on the different numbers of shaping parameters, researchers have proposed several statistical models to characterize the scintillation channel. Among them, two-parameters scintillation models, i.e., Gamma-Gamma and Lognormal-Rician are popularly used, and the latter always performs better than the former, especially under the conditions of weak turbulence, spherical wave, and the receiver with large aperture \cite{Miao20,Churnside87,SONG20124727,Yang15,Andrews2023}. \\
\indent The Lognormal-Rician distribution model is the product of Rician distribution  and  Lognormal distributions, which is described by the coherence parameter $r$ and the  variance $\sigma_z^2$ respectively.  Accurate estimation of channel parameters $\left(r,\sigma_z^2\right)$ plays a vital role in determining the system performance (such as the bit error rate, ergodic capacity and the outage probability) and aiding channel state information estimation in practice \cite{DABIRI2017577}.  However, the complicated integral form makes it less convenient to be handled. The first estimation method for this distribution was developed by the scholars Churnside and Clifford \cite{Churnside87}, which is based on   a physical model of the turbulence-induced scattering. Note that the accuracy of this approach
depends heavily on the scattering physical model, which may
not be readily available. The authors in \cite{SONG20124727} applied the Hansen
two-step generalized method of moments (GMM) method
to estimate the shaping parameters. The advantages of GMM method can avoid the computation of integral involving Bessel function, but the key drawback of GMM is that it can suffer from large bias and inefficiency in small channel samples. For example, it was found in \cite{SONG20124727} that $10^6$ data samples are required to achieve satisfactory estimation  performance for the Lognormal-Rician distribution. However, the order of  1000 seconds latency induced by the $10^6$ data samples is unacceptable for practical  communication systems. Upon addressing this problem, the other expectation-maximization (EM) estimator was developed in \cite{Yang15}. We note this estimation approach requires the computation of complicated
integrals although it provides good estimation performance  with only $10^3$ data samples. More recently, the  first estimation method that achieves the balance between computational complexity and accuracy, namely the saddlepoint approximation (SAP) estimator was formulated by our previous work \cite{Miao20}. This method requires the computation of the saddlepoints based on the expression  involving Bessel function, which is time-consuming to
accomplish from a hardware implementation point of view.    \\
\indent In this paper, we propose a novel and efficient parameter estimation for the
Lognormal-Rician turbulence model based on $k$-Nearest Neighbor ($k$NN) and data
generation method. In contrast to other estimators mentioned above model, we demonstrate that this method avoids the  computations of both the integral and Bessel function while maintaining the estimation accuracy,  thus achieving the best tradeoff between the computation complexity and the accuracy. Specifically, the simulation results indicate that the mean squared error (MSE) of variance of the Lognormal
modulation factor $z$, i.e., $\sigma_z^2$ by our method outperforms  that of the SAP estimator in some  channel scenarios when combined with the genetic algorithm. Most importantly, we emphasize that this method can be flexibly adapted to different fading models in the field of  wireless communication and free-space optical/quantum communication.\\
\indent For the communication system of interest, it can be assumed that the background noise is suppressed perfectly, and this can be implemented by spatial filtering and adaptive optics. The Lognormal-Rician channel model is the product of Lognormal and Rician distributions,  whose probability density function (PDF) is given by \cite{Miao20,Churnside87,Andrews2023}\footnote{We note  the PDF of Lognormal-Rician channel in \cite{Churnside87} contains an error in the argument of the zero-order modified Bessel function of the first kind $I_0(\cdot)$, and the right form is $2\left[\frac{(1-r) r}{z} I\right]^{1 / 2}$.}
\begin{equation}\label{eq1}
 \begin{aligned}
f(I;r,\sigma_z^2)= & \frac{(1+r) e^{-r}}{\sqrt{2 \pi} \sigma_z} \int_0^{\infty} \frac{d z}{z^2} I_0\left(2\left[\frac{(1+r) r}{z} I\right]^{1 / 2}\right) \\
& \times \exp \left(-\frac{1+r}{z} I-\frac{1}{2 \sigma_z^2}\left(\ln z+\frac{1}{2} \sigma_z^2\right)^2\right)
\end{aligned}
\end{equation}
under  intensity normalization condition, where in (\ref{eq1}), $z,  I_0$ represent the Lognormal random variable and the zero-order modified Bessel function
of the first kind respectively. The relationships between empirical parameters $r,\sigma_z^2$ and the  physical characteristics of atmospheric conditions, such as Rytov variance and atmospheric refractive index structure constants, can be seen in \cite{Churnside87}. Similar to \cite{Churnside87,SONG20124727,Yang15}, we consider the channel conditions for different combinations of $r$ and $\sigma_z^2$  to explore the effectiveness of the proposed estimator in general scenarios.  \\
\indent \textbf{The $k$NN Approximation}. To avoid the computations of integral and Bessel function in (\ref{eq1}), we  employ a nonparametric density estimation approaches, namely $k$NN, which is  widely used in statistics and machine learning. According to  \cite{Nonparametric65}, the $k$NN method estimates the density value at point $x$ based on the distance between $x$ and its $k$-th nearest neighbor, and it can be flexibly adapted to any continuous PDF. Compared with  other two nonparametric density estimation approaches, i.e., kernel density estimation and histogram density estimation, this method can be computed very efficiently using kd-tree algorithm when a huge amount of data is considered \cite{Bishop2006}.  \\
\indent For $M$ channel   random samples $\bold{I} = \{I[1],I[2],\cdots,I[M]\}$ generated according to specific shaping parameters, we can construct the $k$NN density estimator at $I[n]$ as \cite{Wang09}
\begin{equation}\label{eq2}
  \hat{p}_k\left(I[n]\right) = \frac{k}{M-1}\frac{1}{c_1(d)\rho_k^d\left(n\right)}
\end{equation}
where $d$ represents the data dimension of $I[n]$, and it is 1 for the considered Lognormal-Rician turbulence channel. $\rho_k\left(n\right)$ denotes the distance between $I[n]$ to its $k$NN in $\{I[j]\}_{j \neq n}$, and $c_1(d)$ is the volume of the unit ball, which is given by
\begin{equation}\label{eq3}
  c_1(d)=\frac{\pi^{d / 2}}{\Gamma(d / 2+1)}.
\end{equation}
\indent Similar to the SAP method,  a normalized factor $c$ can be introduced to renormalize the derived approximate density in (\ref{eq2}),  which is expressed as
\begin{equation}\label{eq4}
  c \approx \frac{1}{\int_{C[1]}^{C[M]} \text { Interpolation }\left(\hat{p}_k(C)\right) d C}
\end{equation}
where the  sequence $\bold{C} = \{C[1],C[2],\cdots,C[M]\}$ is obtained by  sorting the sequence $\bold{I}$ in ascending order, and the $Interpolation$  is adopted for any two adjacent discrete values in $\bold{C}$. For ease of calculation,  a linear interpolation  is assumed. Moreover, it is important to emphasize that the parameter $k$ plays a crucial  role in approximation: a small $k$  leads to a lower bias and a higher variance, and a larger $k$ contributes to decreasing the variance while still guaranteeing a small bias when the samples sizes are large enough, as discussed in \cite{Silverman86}. \\
\indent The procedure of $k$NN approximation for given data samples $\textbf{I}$ is summarized in Algorithm~\ref{alg1}.
\begin{algorithm}
\caption{$k$NN approximation}\label{alg1}
\begin{algorithmic}[1]
\Procedure{KNNApproximation}{$\textbf{I},M,k$}
\State $\textbf{C}\gets \text{AscendingOrderSort}\left(\textbf{I}\right)$
\State $n = 1$
\While{$n\leq M$}
\State Applying (\ref{eq2}) to evaluate the density at $C[n]$
\State $n = n + 1$
\EndWhile\label{euclidendwhile}
\State Applying (\ref{eq4}) to evaluate the normalized factor $c$
\State \textbf{return $c$, $\mathbf{\hat{p}_k}= \{\hat{p}_k\left(C[1]\right),\cdots,\hat{p}_k\left(C[M]\right)\}$}
\EndProcedure
\end{algorithmic}
\end{algorithm} \\
\indent Next, we employ the Kolmogorov-Smirnov (KS) goodness-of-fit statistical tools to investigate the validity of $k$NN approximation and show  the significance of the choice of $k$. According to \cite{papoulis02}, the KS
goodness-of-fit tests measure the maximum value of the absolute
difference between the empirical CDF, $F_{\hat{I}}(\lambda)$,  and the approximate CDF, $F_I(\lambda)$. Thus, the KS test statistic is defined as
\begin{equation}\label{eq5}
 T \triangleq \max \left|F_I(\lambda)-F_{\hat{I}}(\lambda)\right|
\end{equation}
where in (\ref{eq5}), the approximate CDF, $F_{\hat{I}}(\lambda)$ is obtained as
\begin{equation}\label{eq6}
F_{\hat{I}}(\lambda) = c\int_{0}^{\lambda}\text { Interpolation }\left(\hat{p}_k(C)\right)dC.
\end{equation}
\begin{figure}[thbp!]
    \centering
    \begin{tabular}{@{\extracolsep{\fill}}c@{}c@{\extracolsep{\fill}}}
            \includegraphics[width=0.5\linewidth]{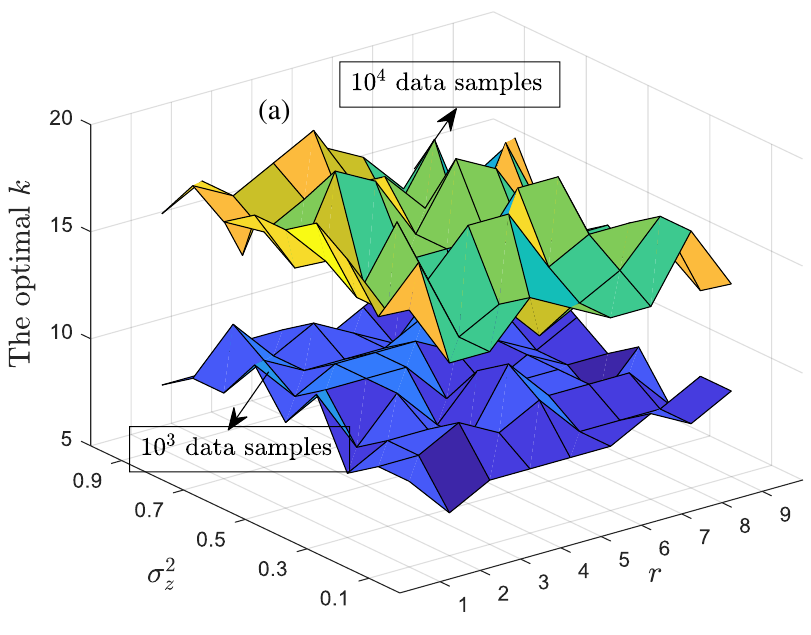} &
            \includegraphics[width=0.5\linewidth]{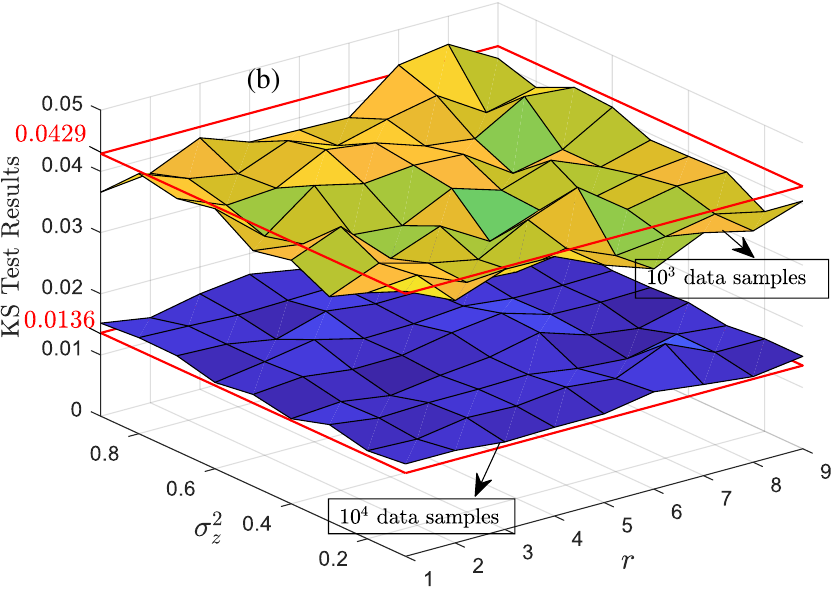} \\
    \end{tabular}
    \caption{The optimal $k$ and KS test results  for different channel conditions. (a)The optimal $k$, (b)KS  test results when $k = 2$.}
    \label{fig1}
 \end{figure}
\indent In Fig.~\ref{fig1}(a), we present the optimal KS test results under   typical channel conditions. Note that the results for each pair of parameters ($r, \sigma_z^2$)  are obtained by averaging the results of 100 simulation runs for each integer $k$, and keeping the minimum value. We find it surprising that the optimal $k$ for $10^3, 10^4$ data samples  are around 8 and  15 for  the considered channel conditions. \\
%\indent  The critical values $T_{max}$ are $0.0429, 0.0136$ respectively for   $10^3, 10^4$ data samples when a typical significance level $\alpha = 5\%$ is considered. It can be clearly observed from the Fig.~\ref{fig1} that the KS test results for $10^3, 10^4$ are less than the corresponding thresholds, indicating an efficient approximation can be achieved by using the $k$NN method with optimal $k$.\\
\indent In Fig.~\ref{fig1}(b), we present the  KS test results under  different channel conditions when $k = 2$. Given a significance level $\alpha = 5\%,  $the critical values $T_{max} = 0.0429, 0.0136$  colored in red  for   $10^3, 10^4$ data samples are also included as a benchmark \cite{papoulis02}. It can be observed that the KS test results are near the critical values  for both the data samples in this case.
%\begin{figure}[h]
%\centering
%{\includegraphics[width=2in]{PDF/Fig5.pdf}}
%\caption{KS goodness-of-fit test results between the CDF of $\mathbf{\hat{p}_k}$ and the CDF of empirical distribution under different channel conditions when $k = 2$.}
%\label{fig5}
%\end{figure}\\

\begin{algorithm}
\caption{The approximate computation of LLF}\label{alg2}
\begin{algorithmic}[1]
\Procedure{LLFApproximation}{$\textbf{C}, L, \hat{k}, \hat{r}, \hat{\sigma_z^2}$}
\State Generate $L$  samples $\textbf{Z} = \{Z[1],\cdots,Z[L]\}$, \Comment{each sample in \textbf{Z}  follows the Lognormal-Rician distribution with shaping parameters $\left(\hat{r},\hat{\sigma_z^2}\right)$}
\State $\textbf{T} \gets \text{AscendingOrderSort}\left(\textbf{Z}\right)$
\State $c, \mathbf{\hat{p}_k} \gets \text{KNNAPPROXIMATION}\left(\textbf{T},L,\hat{k}\right)$
\State $\mathbf{CTemp} \gets \text{Select samples from } \mathbf{C} \text{ that are in} \left[T[1],T[M]\right]$
\State $MTemp \gets \text{Length(CTemp)}$
\While{$n\leq MTemp$}
\State Find the interval $[T[j], T[j+1]]$ that contains  the  $  CTemp[n] $
\State Evaluate the density at CTemp[n], $\hat{p}_k\left(CTemp[n]\right)$ according to $\left(T[j],\hat{p}_k\left(T[j]\right)\right)$, $\left(T[j+1],\hat{p}_k\left(T[j+1]\right)\right)$ and a linear interpolation between them.
\State $n = n + 1$
\EndWhile\label{euclidendwhile}
\State Evaluate the $\text{LLF} = \frac{1}{MTemp}\sum  \limits_{n=1}^{MTemp}ln\left(\hat{p}_k\left(CTemp[n]\right)\right) + c $
\State \textbf{return LLF}
\EndProcedure
\end{algorithmic}
\end{algorithm}
\indent \textbf{Construction of a new estimator}. To reveal the importance of the $k$NN approximation, a novel and efficient estimator can be developed by combining it with the data generation method.\\
\indent In Algorithm 2,  we firstly  show the process to compute the log-likelihood function (LLF) of channel random samples $\mathbf{C}$ approximately under   Lognormal-Rician distribution with shaping parameters (${r}, {\sigma_z^2}$). Following the  methodology of maximum likelihood estimation and SAP estimation,  we expect that the optimum estimated shaping parameters that  maximizing the LLF is  near  the actual values  when the optimal $k$ is achieved.  Thus,  the objective function can be  formulated   as
\begin{equation}\label{eq7}
  \max _{{k}, {r}, {\sigma_z^2}} L L F\left({k}, {r}, {\sigma_z^2}\right).
\end{equation}
It must be emphasized that the channel  samples $M$ depends on the turbulence channel coherence time, and can not be large for the practical systems. However, by using a digital signal processing (DSP) chip with mass memory at the receiving end, the  number of generated  samples $L = length(\textbf{Z})$ can be large enough to achieve a highly accurate approximation. According to (\ref{eq7}), we can find an  additional parameter $k$ also needs to be estimated, which is the same as the SAP estimators.\\
\indent Furthermore,  given the channel samples,  the obtained LLF is not constant as it is approximately evaluated by the data generation method, and the fewer channel samples or generated  samples always lead to a large variance, which  is not conducive to accurate estimation of shaping parameters.  Thus, a new estimator can be finally   formulated as
\begin{equation}\label{eq8}
  \max _{\hat{k}, \hat{r}, \hat{\sigma_z^2}} \frac{1}{N_{LLF}} \sum\limits_{n=1}^{N_{LLF}} L L F_{n}\left(\hat{k}, \hat{r}, \hat{\sigma_z^2}\right)
\end{equation}
by  taking the average of multiple
runs, where $N_{LLF}$  represents the number of calculations for  LLF.

\begin{figure}[htbp]
    \centering
    \begin{tabular}{@{\extracolsep{\fill}}c@{}c@{\extracolsep{\fill}}}
            \includegraphics[width=0.4\linewidth]{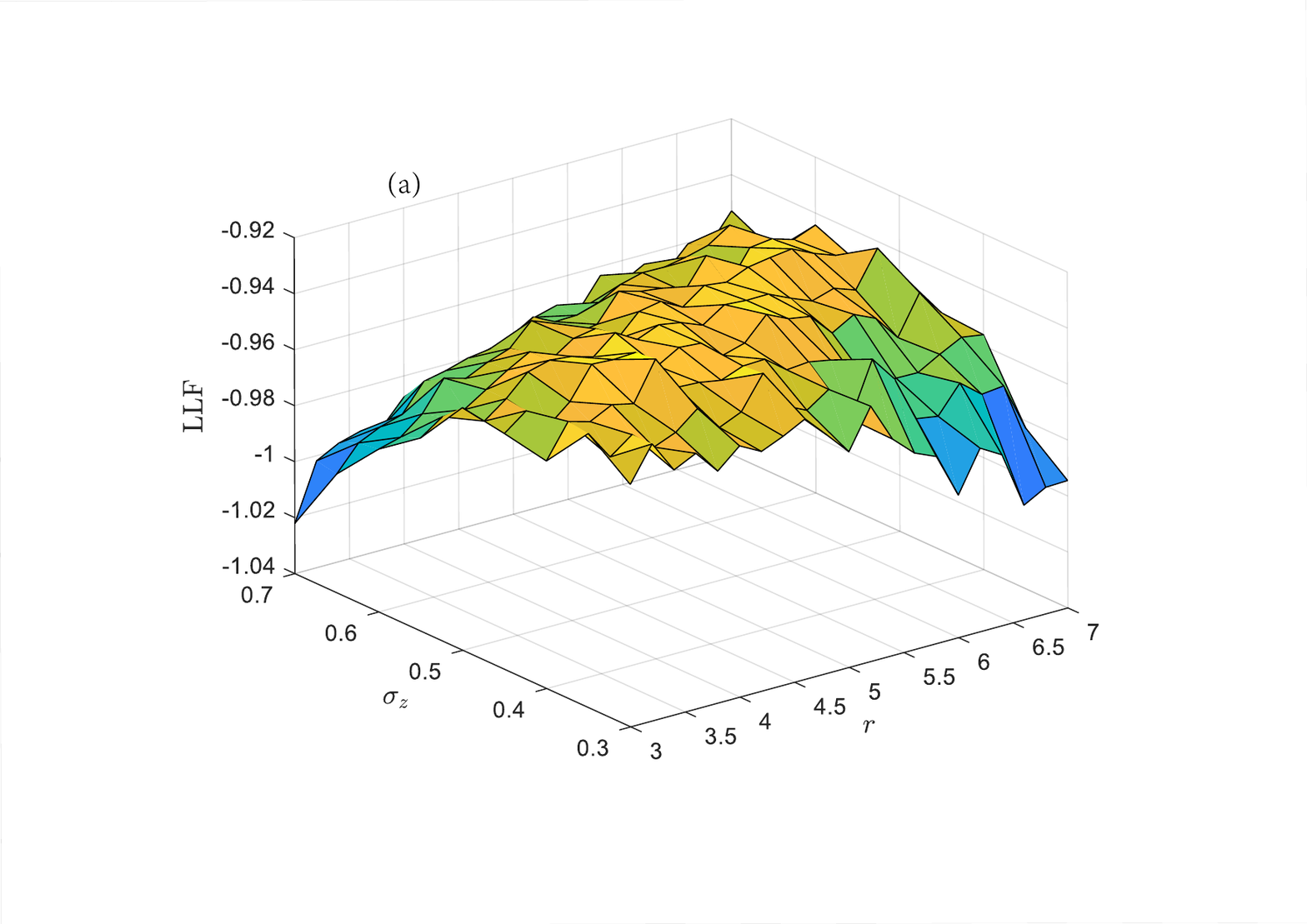} &
            \includegraphics[width=0.4\linewidth]{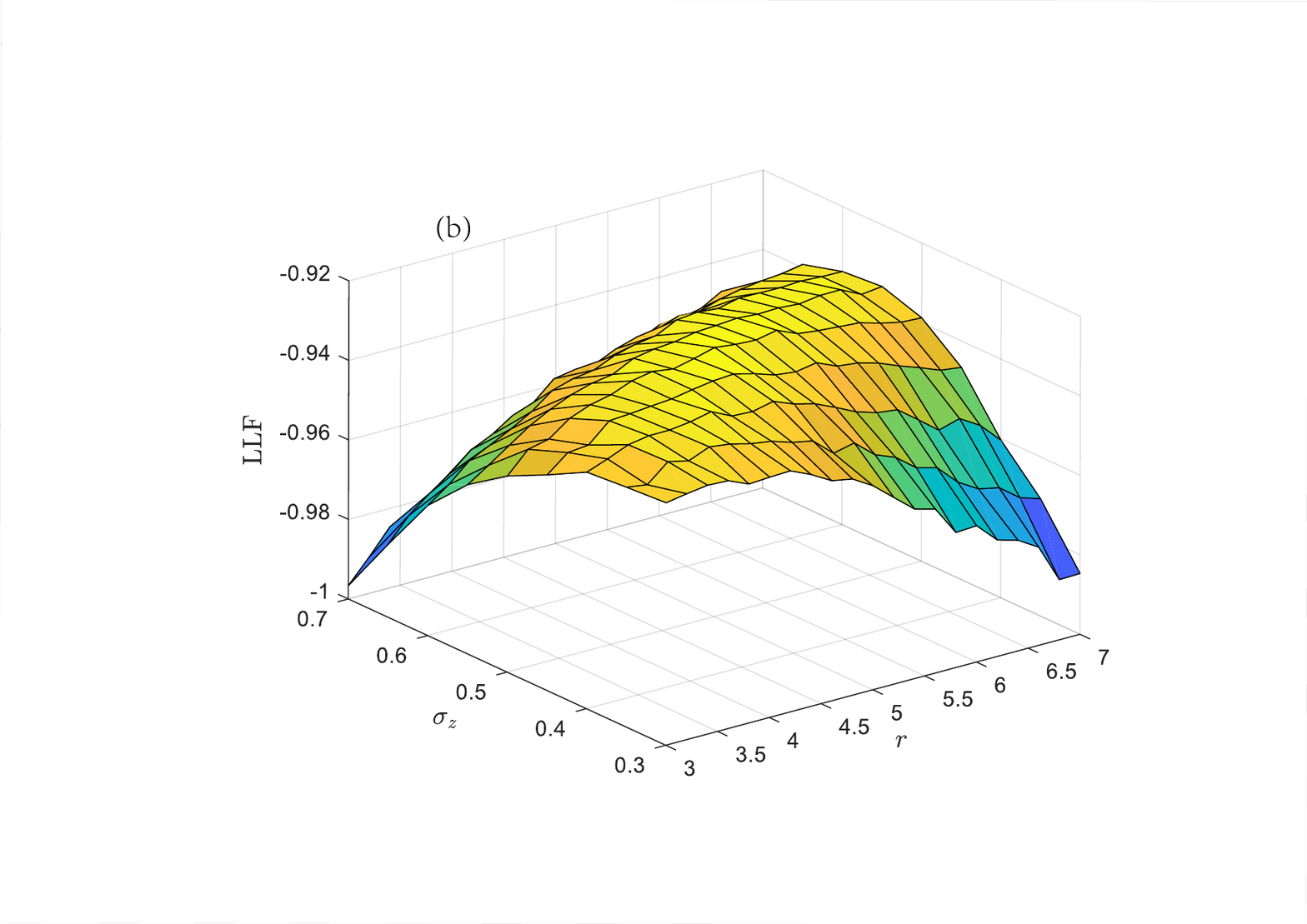}\\
            \includegraphics[width=0.4\linewidth]{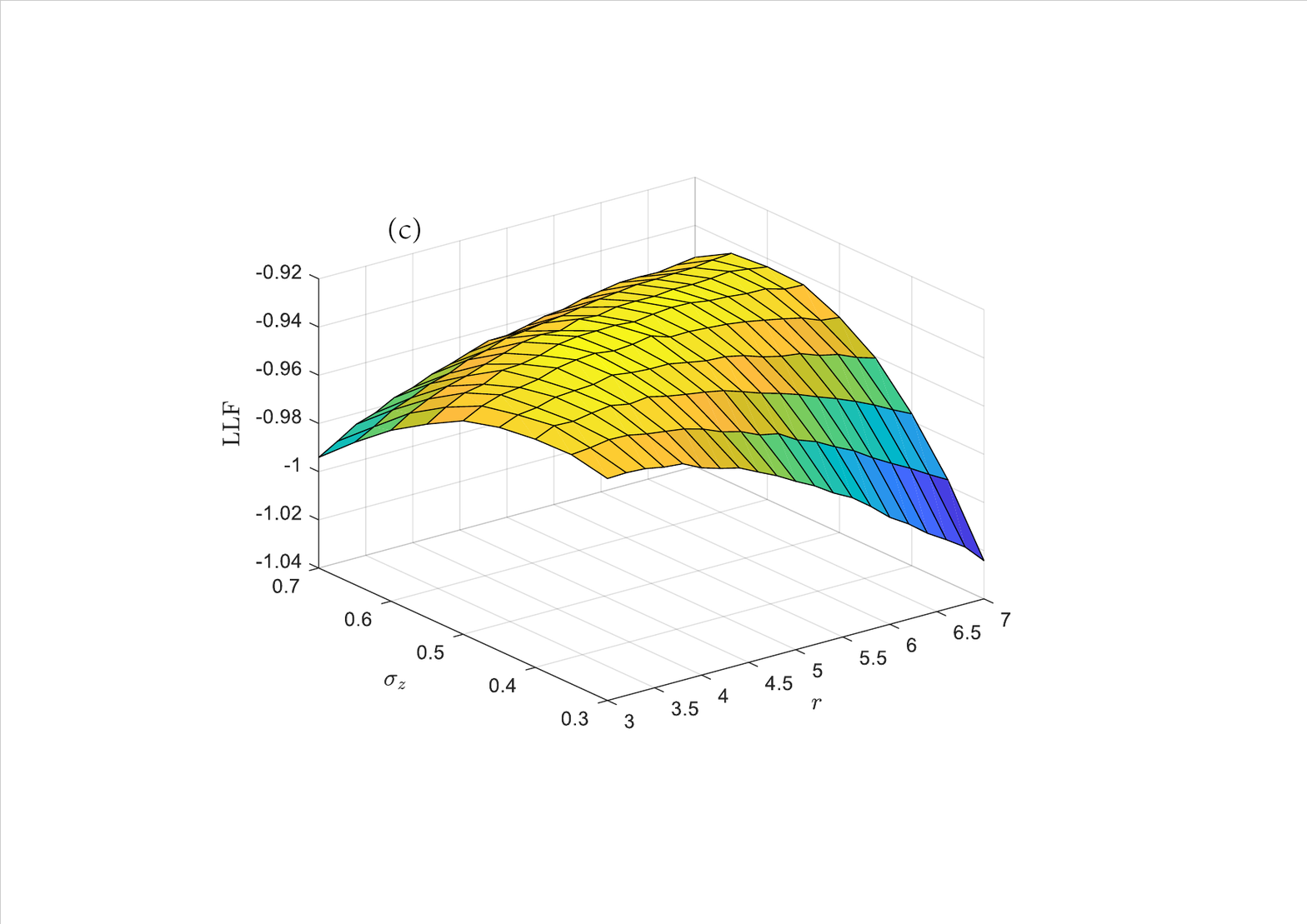}&
            \includegraphics[width=0.4\linewidth]{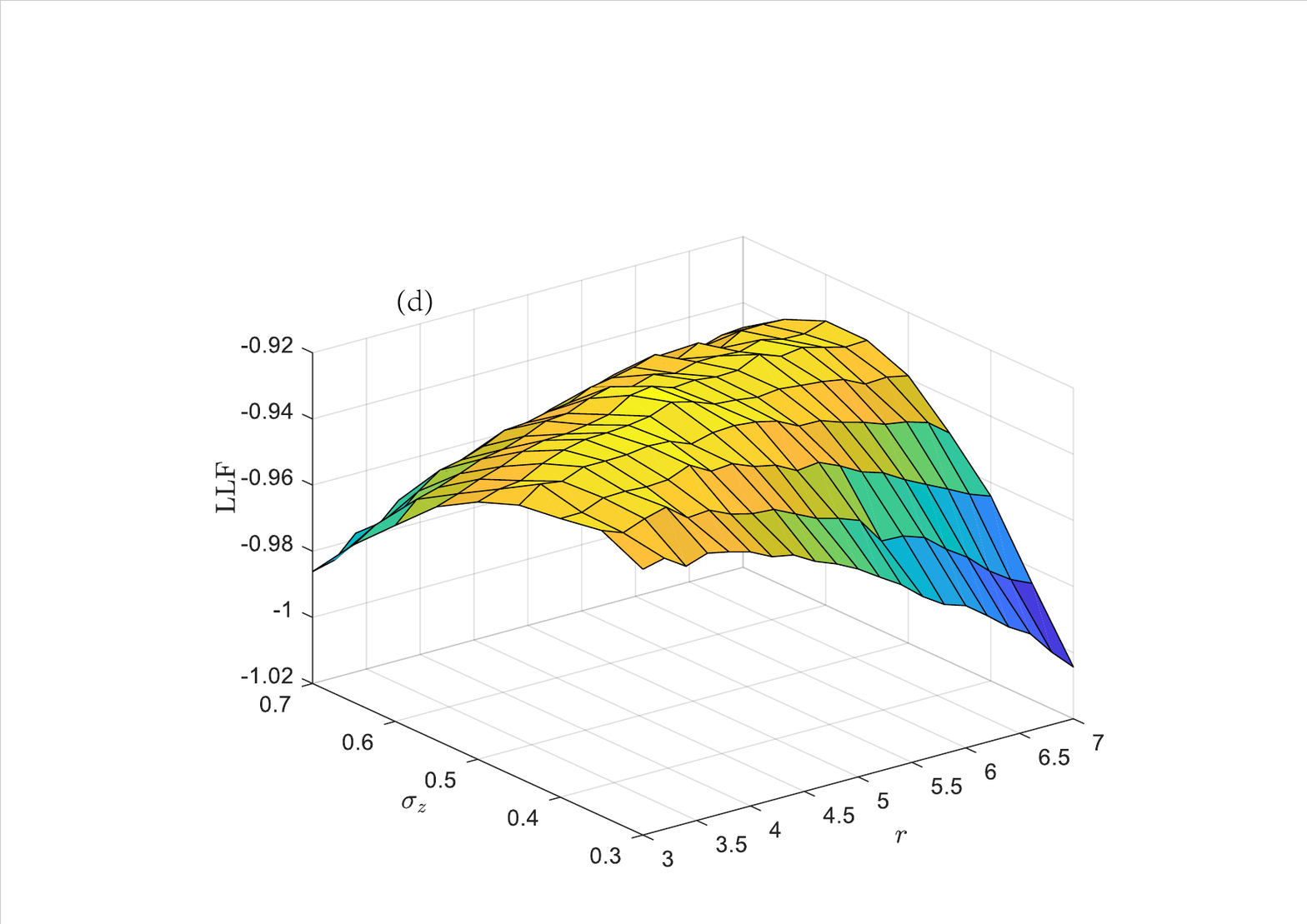}\\
    \end{tabular}
    \caption{LLFs under different channel conditions. (a)$M = 10^4$, $L = 10^4, N_{LLF}=1$, ${r}^{*}= 3.8, {\sigma_z^2}^{*} = 0.16$, (b)$M = 10^4$, $L = 10^4, N_{LLF}=20$, ${r}^{*} = 5.6, {\sigma_z^2}^{*} = 0.25$, (c)$M = 10^4$, $L = 10^6, N_{LLF}=20$, ${r}^{*} = 5, {\sigma_z^2}^{*} = 0.25$, (d) $M = 10^3$, $L = 10^6, N_{LLF}=50$,  ${r}^{*} = 4.8, {\sigma_z^2}^{*}= 0.25$.}
    \label{fig2}
 \end{figure}
\indent In Fig.~\ref{fig2}, we present some numerical results to illustrate that solving the objective function in (\ref{eq8}) can provide a reasonable estimate for the actual values. In these figures, the  values $r, \sigma_z^2$, and $k$ are set to be $5, 0.25$, and $15$ respectively, and all the LLFs are evaluated under the same channel samples. The optimal estimate of the parameters is the point that the LLF is maximized, which is represented by ${r}^{*}, {\sigma_z^2}^{*}$. \\
\indent In Fig.~\ref{fig2}(a), the LLFs are shown under the conditions $M = L = 10^4, N_{LLF} = 1$, and it can be seen that the LLFs around the optimal estimate fluctuate severely, which results to a worse estimate. For this case, we should  employ some intelligent algorithms, like  genetic algorithm (GA), and simulated annealing  since conventional gradient-based methods are not feasible \cite{Pham98}. Then, according to Fig.~\ref{fig2}(b)-Fig.~\ref{fig2}(c), with the increasing   number of calculations for the LLF and the generated  samples, it can be clearly found that the LLFs around the actual values become more stable and the surface becomes more smooth,  which enables us to solve the (\ref{eq8}) by using the gradient-based methods. Specifically, the estimates in Fig.~\ref{fig2}(c) are equivalent to the actual values, which shows that more generated  samples lead to a better estimation in this situation. Moreover,  by averaging 50 LLFs for each pair shaping parameters, the optimal estimate is $r^*=4.8, {\sigma_z^2}^*=0.25$, which indicates that the proposed method also shows an excellent performance with fewer channel samples $M =10^3$, as shown in Fig.~\ref{fig2}(d).\\
\indent \textbf{Simulation environment and MSE performance.} We investigate the proposed estimator performance by using the MSE  of $\hat{\bm{\theta}}$, which is defined as $\text{MSE}\left[{\bm{\theta}}\right] = \operatorname{var}[\widehat{\boldsymbol{\theta}}]+(\mathbb{E}[\widehat{\boldsymbol{\theta}}]-{\theta})^2$ with $\mathbb{E}$ and $\theta$ denoting the expectation and the actual value respectively. The Lognormal and Rician random variates are generated using the \emph{random} function in MATLAB and 35 trials are employed to calculate the MSE performance of the
estimator. We note the  number of calculations for the LLF is 50 when using the GD algorithm  while it is only 1 when using the GA. In addition, the initial estimates of shaping parameters $\hat{r}^{(0)}$ and ${\hat{\sigma_z^2}}^{(0)}$  are obtained according to (5) and ${\hat{\sigma_z^2}}^{(0)}=-\frac{2}{K} \sum_{l=0}^{K-1} \ln I[l]$ in \cite{Yang15}.  With respect to the GA, we note  it is implemented by using the \emph{ga} function in MATLAB's global optimization toolbox. The population size and selection strategy  for the next generation in \emph{ga}  are  set to be 100 and "tournament selection".
\begin{figure}[!h]
\centering
{\includegraphics[width=2.4in]{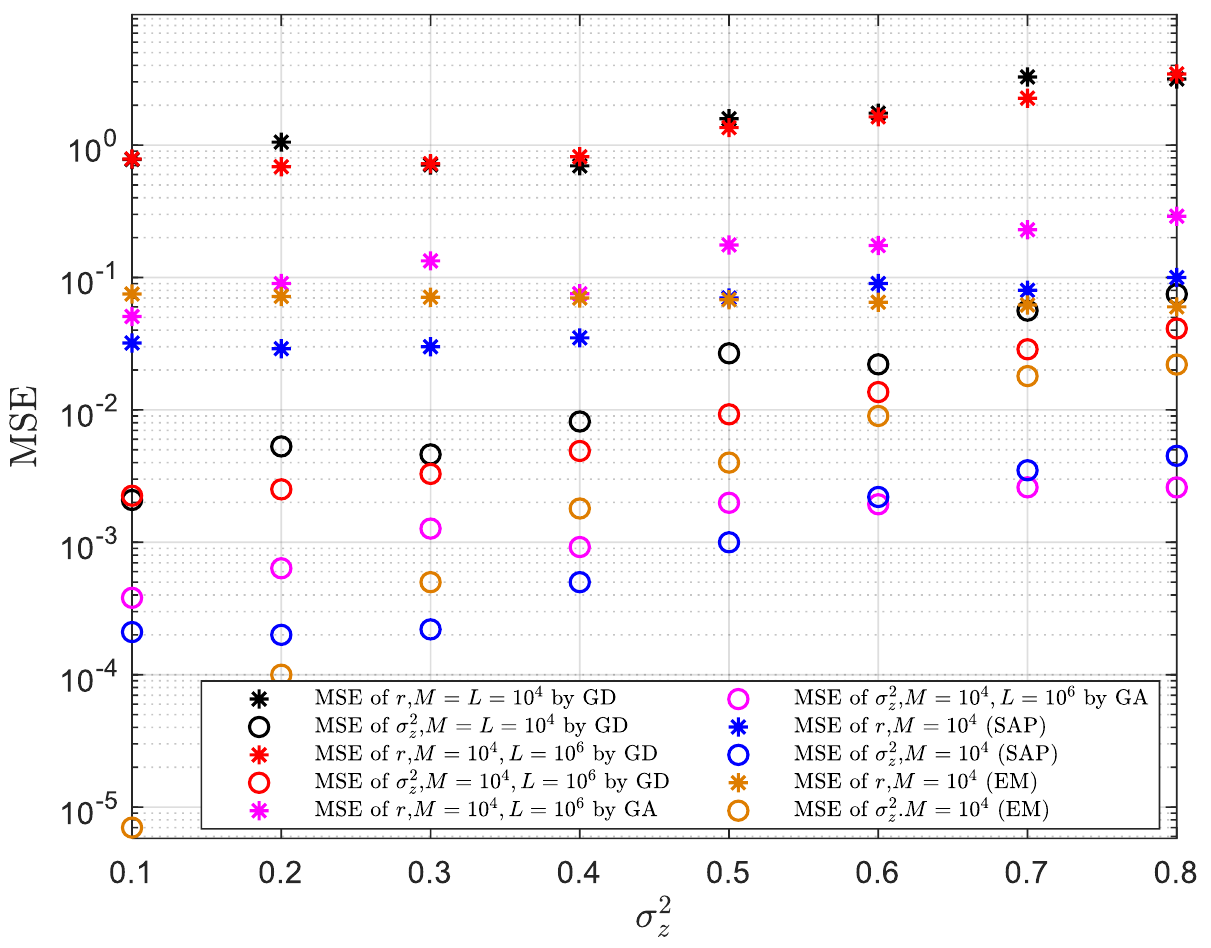}}
\caption{MSE performance of the  estimators under different Lognormal-Rician channel conditions, where $r  =4$. }
\label{fig6}
\end{figure}

\begin{figure}[!h]
\centering
{\includegraphics[width=2.4in]{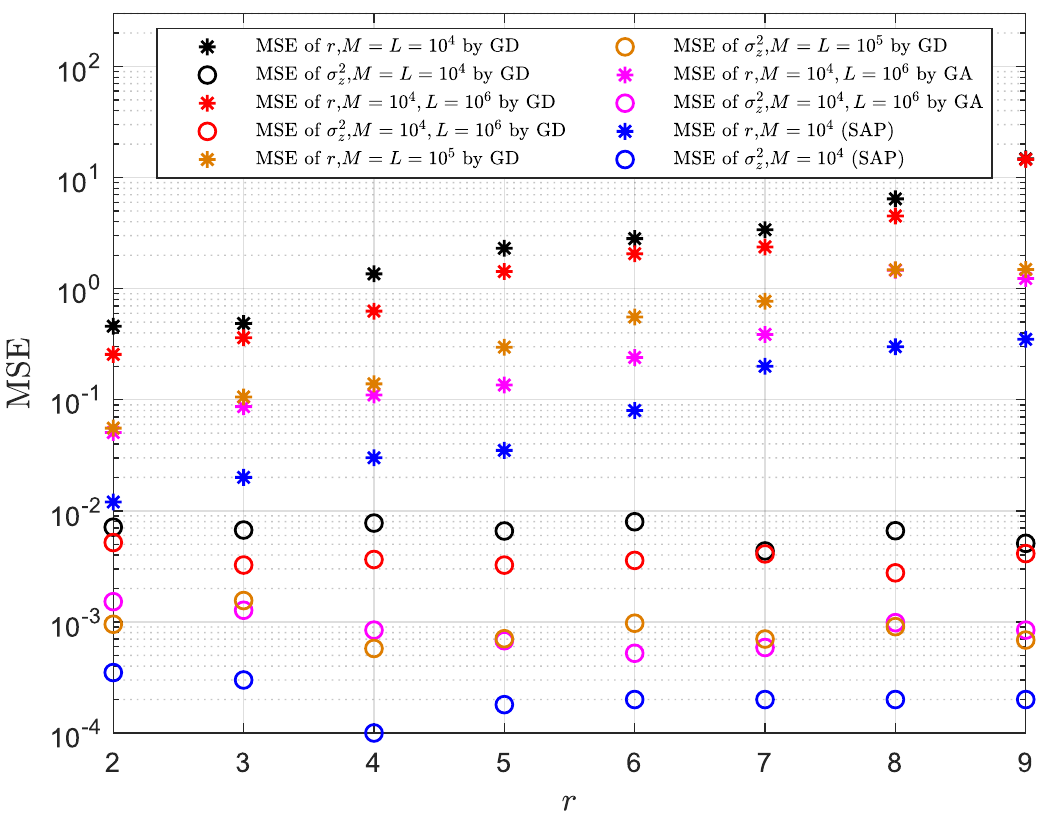}}
\caption{MSE performance of the  estimators under different Lognormal-Rician channel conditions, where $\sigma_z^2 = 0.25$. }
\label{fig7}
\end{figure}
\indent Fig.~\ref{fig6} shows the simulated MSE performance of $r, \sigma_z^2$ when $r = 4$. The  performance of  SAP and EM estimators is also included for comparison. As can be seen from this figure, increasing the number of generated  samples by two orders of magnitude does not lead to a significant improvement in  estimation performance when solving the optimization problem by the GD algorithm. However, the MSE performance with the GD algorithm can be improved by nearly five times for both the shaping parameters with the GA. In addition,  compared with the SAP  and EM estimators, we can find the proposed estimator with the GA achieves the best estimation performance for $\sigma_z^2$ when $\sigma_z^2 \geq 0.6$ while the estimation performance for $r$ is about two times worse than that obtained by the EM estimator. \\
\indent In Fig.~\ref{fig7}, we present the simulated MSE performance for shaping parameters when $\sigma_z^2 = 0.25$. We can also draw a conclusion that increasing the number of generated  samples  does not lead to a significant improvement in  estimation performance with the GD algorithm. In addition, it can be observed that the MSE performance of $\sigma_z^2$ is insensitive to the value of $r$, but it is sensitive to the value of $\sigma_z^2$ when comparing the curves between Fig.~\ref{fig6} and Fig.~\ref{fig7}, and this is consistent with the results presented  in \cite{Miao20} and \cite{Yang15}.  Interestingly,  the estimation performance for $M = 10^4, L = 10^6$ by the GD algorithm is nearly the same as that for $M = L = 10^5$ by  GA, and they are slightly worse than that of the SAP estimator. \\
\indent \textbf{Conclusion.} In conclusion, a novel and efficient parameter estimation approach  by utilizing the $k$NN and data generation method for the Lognormal-Rician turbulence channel is proposed.  The validity of the $k$NN approximation is investigated by the KS statistical tool, and  the optimal $k$ can achieve an efficient approximation under different channel conditions. The LLFs numerical results indicate that   maximizing the   objective function gives a reasonable estimate for the actual values.   The MSE simulation results demonstrate that the performance of the proposed estimator with the GA approximates to that of the SAP and EM estimators, which achieves the best tradeoff between
the computation complexity and the accuracy. Finally, it is worth mentioning that our proposed estimator can be flexibly adapted to different fading models in the field of wireless communication and free-space optical/quantum communication.
\begin{backmatter}
\bmsection{Funding} National Natural Science Foundation
of China (Grant No.61871347,  No.12405026), Zhejiang Provincial Natural Science Foundation of China (Grant No. LQN25F010019), and Natural science Foundation of Hangzhou (Grant No. 2024SZRYBA050001).
\bmsection{Acknowledgement} The authors would like to thank the  supercomputing Center of Hangzhou City University for the  simulation support.

\bmsection{Disclosures} The authors declare no conflicts of interest.

\bmsection{Data Availability Statement} Data underlying the results presented in this paper are
not publicly available at this time but may be obtained from the authors upon
reasonable request.

\end{backmatter}

% Bibliography
\bibliography{sample}

% Full bibliography added automatically for Optics Letters submissions; the following line will simply be ignored if submitting to other journals.
% Note that this extra page will not count against page length
\bibliographyfullrefs{sample}

%Manual citation list
%\begin{thebibliography}{1}
%\bibitem{Zhang:14}
%Y.~Zhang, S.~Qiao, L.~Sun, Q.~W. Shi, W.~Huang, %L.~Li, and Z.~Yang,
 % \enquote{Photoinduced active terahertz metamaterials with nanostructured
  %vanadium dioxide film deposited by sol-gel method,} Opt. Express \textbf{22},
  %11070--11078 (2014).
%\end{thebibliography}

% Please include bios and photos of all authors for aop articles
\ifthenelse{\equal{\journalref}{aop}}{%
\section*{Author Biographies}
\begingroup
\setlength\intextsep{0pt}
\begin{minipage}[t][6.3cm][t]{1.0\textwidth} % Adjust height [6.3cm] as required for separation of bio photos.
  \begin{wrapfigure}{L}{0.25\textwidth}
    \includegraphics[width=0.25\textwidth]{john_smith.eps}
  \end{wrapfigure}
  \noindent
  {\bfseries John Smith} received his BSc (Mathematics) in 2000 from The University of Maryland. His research interests include lasers and optics.
\end{minipage}
\begin{minipage}{1.0\textwidth}
  \begin{wrapfigure}{L}{0.25\textwidth}
    \includegraphics[width=0.25\textwidth]{alice_smith.eps}
  \end{wrapfigure}
  \noindent
  {\bfseries Alice Smith} also received her BSc (Mathematics) in 2000 from The University of Maryland. Her research interests also include lasers and optics.
\end{minipage}
\endgroup
}{}

\end{document}